\documentclass[12pt]{iopart}
\usepackage{graphicx}
\usepackage{epsfig}

\begin{document}

\title[Numerical simulation of the magnetization of high-temperature superconductors]{Numerical simulation of the magnetization of high-temperature superconductors: 3D finite element method using a single time-step iteration}

\author{  Gregory P. Lousberg$^{1,2}$, Marcel Ausloos$^{3}$, Christophe Geuzaine$^{4}$, Patrick Dular$^{4}$, Philippe Vanderbemden$^{1}$, and Benoit Vanderheyden$^{1}$}
  \address{ $^{1}$ SUPRATECS, Department of Electrical Engineering and Computer Science (B28), University of Li\`ege, Belgium} 
  \address{$^{2}$ FRS-FNRS fellowship}
  \address{$^{3}$ SUPRATECS (B5a), University of Li\`ege, Belgium}
  \address{$^{4}$ ACE, Department of Electrical Engineering and Computer Science (B28), University of Li\`ege, Belgium}
  \ead{ gregory.lousberg@ulg.ac.be}

\begin{abstract}
\linespread{2}
   \selectfont 
 
We make progress towards a 3D finite-element model for the magnetization of a high temperature superconductor (HTS): We suggest a method that takes into account demagnetisation effects and flux creep, while it neglects the effects associated with currents that are not perpendicular to the local magnetic induction. We consider samples that are subjected to a uniform magnetic field varying linearly with time. Their magnetization is calculated by means of a weak formulation in the magnetostatic approximation of the Maxwell equations ($A$-$\phi$ formulation). An implicit method is used for the temporal resolution (Backward Euler scheme) and is solved in the \textit{open source} solver GetDP. Picard iterations are used to deal with the power law conductivity of HTS.  The finite element formulation is validated for an HTS tube with large pinning strength through the comparison with results obtained with other well-established methods. We show that carrying the calculations with a \textit{single} time-step (as opposed to many small time-steps) produce results with excellent accuracy in a drastically reduced simulation time. The numerical method is extended to the study of the trapped magnetization of cylinders that are drilled with different arrays of columnar holes arranged parallel to the cylinder axis. 

\end{abstract}

\submitto{\SUST}


\maketitle
\linespread{2}
   \selectfont 
\section{Introduction}
\label{s:Intro}

Bulk high-temperature superconductors (HTS) become increasingly attractive for being used as efficient magnetic shields~\cite{Sam} or as powerful permanent magnets~\cite{A1,Murakami}. Highly sensitive magnetic measurement systems, such as in a biomagnetic imaging device, need efficient magnetic shields for reducing the effects of the external magnetic disturbances~\cite{MagnShield1,MagnShield2,MagnShield3}. Powerful magnets are required in magnetic bearing systems, where they produce large levitation forces~\cite{A3,A6}, or in rotating machines, where they produce a large torque on the shaft~\cite{A2,A4,A5}.

The performances of HTS trapped field magnets are limited by three main factors: (i) the critical current density in the sample, $J_c$, that determines the maximum trapped magnetic field; (ii) the strength of the mechanical stresses, that arise  from strong Lorentz forces and may result in cracks in the sample; and (iii) the heat exchange rate with the cryogenic fluid, that when too low may lead to significant temperature rises if the sample is subjected to a variable magnetic flux, as it is the case in rotating machines~\cite{TemperatureRise}. 

In order to improve the performances of HTS magnets, attention has recently turned to bulk samples in which an array of parallel columnar holes is drilled along the $c$-axis~\cite{Drilled2,H3,H4}. The hole array enables one to obtain a better oxygen annealing~\cite{H5} --- and therefore to raise $J_c$ ---, to perform a more efficient cooling~\cite{Drilled3}, or to reinforce mechanically the sample by injecting a resin~\cite{Murakami,H4}. On the other hand, the holes also block the current stream lines --- which have to flow around them --- and, as a result, degrade the magnetic properties of the sample. In a previous work~\cite{Drilled}, the Bean critical state model was used for calculating the first magnetization of drilled cylinders of infinite extension, as a function of their hole pattern. It was shown that the penetration of the magnetic flux in a given hole generated a discontinuity in the flux distribution ahead of that hole. The trapped magnetic flux could then be increased by placing the holes on the discontinuity lines (\textit{i.e.} lines where the current flow abruptly changes its direction due to the presence of a hole) of their direct neighbors, as this arrangement limited the perturbation by an individual hole of the overall flux distribution.

This study was applied to samples of infinite extension and thus neglected demagnetisation effects. To pursue the study and compare with experimental results, work is needed to model the three-dimensional distribution of the magnetic flux while taking due account of the presence of the holes, the actual path followed by the current lines, and the resulting demagnetisation effects.

Calculating a three-dimensional magnetic field distribution in HTS is notoriously difficult~\cite{Brandt,Mikitik}. In the limit of infinite pinning strength, the magnetic flux distribution can be described with the concept of the critical state, introduced by Bean~\cite{Bean}. The critical state is characterized by a current density with a constant magnitude that flows perpendicular to the local flux density lines, since the magnetic force exerted on the vortices only depends on that component~\cite{Mikitik}. For a series of geometric configurations with a high level of symmetry, e.g. a cylinder subjected to an applied field with an axial symmetry, the current density is known to be everywhere perpendicular to the magnetic field, and the critical state can be simpled determined~\cite{Bean, Brandt2}. However, for an arbitrary configuration where either the sample or the source of the field has no particular symmetry, the critical state model must be modified in order to properly describe the time evolution of the component of the current density that is parallel to the local magnetic field~\cite{Mikitik}. Moreover, for realistic pinning strengths, flux creep must also be taken into account, particularly at the temperature of liquid nitrogen ($77~\mathrm{K}$)~\cite{fluxcreep,fluxcreep2,Rhyner}. To our knowledge, no model includes yet a proper treatment of both the longitudinal component of the current density and the creep effects.

The purpose of this work is to make progress toward a 3D model for the calculation of the trapped flux in drilled HTS magnets, by taking into account demagnetisation effects and flux creep, while neglecting the more delicate effects associated with longitudinal currents. In practice, we expect such a description to faithfully reproduce the actual flux distribution near the median plane of the sample, since the current lines are expected to lie in the plane and thus to be perpendicular to the local flux lines. Moreover, we believe that the neglect of those effects associated with the longitudinal component of the current density leads nonetheless to a first-order approximation from which \emph{qualitative} conclusions can be drawn regarding the influence of the hole lattice on the magnetic properties of the drilled samples.

In the rest of this paper, HTS are modeled by a power law conductivity $\sigma (\mathbf{E}) $ ~\cite{Rhyner} assuming the form
\begin{equation}
\mathbf{J}=\sigma(\mathbf{E})\mathbf{E}=\frac{J_c}{E_c^{1/n}}\left(|\mathbf{E}|\right)^\frac{1-n}{n}\mathbf{E}\label{E(J)},
\end{equation}
where $E_c$ is the critical electric field and $J_c$ is the critical current density. The critical exponent, $n$, is related to the pinning strength in the material and is assumed to be independent of the magnetic field. When $n=1$, we recover the constitutive law of an ohmic material. In the opposite limit of infinite pinning strength, $n \to \infty$, the power law model is asymptotically equivalent to the Bean model~\cite{Prigozhin,Yin}. Several numerical methods are available to solve for the magnetic field penetration in HTS with a power law conductivity: finite-difference approximation in cylinders~\cite{Berger}, Green's function approach in cylinders~\cite{Brandt} or in tubes~\cite{Sam}, and finite element method (FEM) with so-called $A$-$\phi$ formulations~\cite{Ruiz,Grilli}, $T$-$\Phi$ formulations~\cite{Amemiya,Tixador}, or unconstrained $H$-formulations~\cite{Hong,Pecher}.
In each of these methods, the choice of the time-step is crucial since it governs the convergence rate and the total calculation time, which can become excessively long on a 3D mesh when $n$ is large~\cite{Grilli,Hong,Roy}. 

To our knowledge, in the FEM suggested so far, the computation time-step was chosen much smaller than the timescale characterising the simulated external excitation. Such a choice can however be largely improved in the (present) case of an excitation varying linearly with time. Our argumentation is two-fold. First, from the point of view of the physics involved, one knows that the vortex motion and flux creep are strongly reduced as the pinning strength increases. Thus, for large $n$, the motion of vortices can only be induced by applying an external flux variation, so that the time behaviour of the magnetic response is expected to be mainly dictated by the excitation sweep rate, not by creep effects.  The second part of our argumentation stems from the numerics involved. We solve a time-differential equation of the form $\partial u / \partial t = g(u)$ with the backward Euler scheme~\cite{NumericalMethod}. The temporal derivative at time $t$ is approximated at first-order, yielding the implicit equation
\begin{equation}
\frac{u_t-u_{t-\Delta t}}{\Delta t}=g(u_t).
\end{equation}
Such a scheme has been shown to yield a truncature error proportional to the second time derivative, $e_t \approx \partial^2 u/\partial t^2\, \Delta t + O(\Delta t^2)$~\cite{NumericalMethod}. Again, in the limit of large pinning strength and with an external field applied as a ramp, we expect the second time derivative of the magnetic response to be small, as its timescale is dictated by that of the excitation, which varies here linearly with time. These arguments suggest that larger $\Delta t$ can be used, and in the extreme case, a single time-step might be used.

This paper addresses the questions of the accuracy and the convergence of a single time-step method that is suitable for a 3D model of HTS. For this purpose, we use a finite-element formulation implemented in the \textit{open source} numerical solver GetDP~\cite{GetDP,GetDP1}. The rest of this manuscript is organised as follows: in Section~\ref{s:formulation}, we describe and motivate the choice of an $A$-$\phi$ formulation. In Section~\ref{s:computation}, we describe the implementation of this formulation into GetDP, and validate it in Section~\ref{s:tube}, where comparisons are made with the Bean model in the case of an HTS tube with an infinite height (2D geometry), and with the Green's function method~\cite{Brandt} in the cases of a tube of finite height (3D geometry). In particular, we analyse the validity of the single time-step method as a function of the value of the critical exponent $n$ and the ramping rate. In section~\ref{s:drilled}, we apply the FEM for calculating the trapped magnetic flux density in drilled HTS cylinders with a finite height, for four different periodical arrangements of the columnar holes. We then conclude in Section~\ref{s:Conclusion}.

\section{Finite element $\mathbf{A}-\phi$ formulation}
\label{s:formulation}

The description of the magnetic field penetration in HTS is based on magneto-quasistatic approximation of the Maxwell equations~\cite{Jackson}. The HTS conductivity is given by Equation~(\ref{E(J)}) and the lower critical field, $H_{c1}$, is neglected against the applied field, so that the material follows the constitutive law, $\mathbf{B}=\mu_0\mathbf{H}$. We introduce the vector potential $\mathbf{A}$ and the scalar potential $\phi$, through 
\begin{eqnarray}
\mathbf{B} &=&\mathbf{B}_{\mathrm{self}}+B_a(t)\mathbf{e_z}=\nabla\times\mathbf{A}+\nabla\times\mathbf{A}_a,\\
\mathbf{E}&=&-\frac{\partial\mathbf{A}}{\partial   t}-\frac{\partial\mathbf{A}_a}{\partial t}-\nabla\phi,
\end{eqnarray}
where the magnetic flux density is split into two contributions: the uniform applied magnetic flux density, $B_a(t)\,\mathbf{e_z}$, which points along the $z$-axis and varies linearly with time as $B_a(t)=\dot{B}_a\,t$, and the reaction magnetic flux density, $\mathbf{B}_{\mathrm{self}}$, which is produced by the eddy currents induced in the HTS. In cylindrical coordinates, the vector potential corresponding to the uniform applied magnetic flux density is given by $\mathbf{A}_a=-r/2B_a(t)\,\mathbf{e}_\theta$ and we have $\partial\mathbf{A}_a/\partial t=-r/2\dot{B_a}\,\mathbf{e}_\theta$.
The introduction of the potentials $\mathbf{A}$ and $\phi$ into the magneto-quasistatic Maxwell equations leads to two coupled equations: 
\begin{equation}
\nabla\times\nabla\times\mathbf{A} =\mu_0\sigma(\mathbf{A},\phi)\left(-\dot{\mathbf{A}}-\dot{\mathbf{A}_a}-\nabla\phi\right)\label{Max1},
\end{equation}
\begin{equation}
\nabla\cdot\left\{\sigma(\mathbf{A},\phi)\left(-\dot{\mathbf{A}}-\dot{\mathbf{A}_a}-\nabla\phi\right)\right\}=0\label{Max2},
\end{equation}
where the electrical conductivity $\sigma$ is calculated from the power law~(\ref{E(J)}), as $\sigma=J_c/E_c^{(1/n)}|\mathbf{E}|^{(1-n)/n}$. These equations are sufficient to describe the electromagnetic behavior of HTS in the $A$-$\phi$ formulation~\cite{Ruiz,Bossavit}. The choice of this particular formulation is motivated by the fact that it produces a strong knowledge of the magnetic flux density, which is the quantity that is directly available in experiment. The Dirichlet boundary conditions on $\mathbf{A}$ and $\phi$ are imposed on the outer surface of a circular shell (in 2D geometry) or a spherical shell (in 3D geometry), whose external surface in both cases is sent to infinity by a Jacobian transformation~\cite{Jacobian}. At infinity, we set $\mathbf{A}=0$ and $\phi=0$.

Equations~(\ref{Max1}-\ref{Max2}) are solved by the Galerkin residual minimization method, which yields
\begin{equation}
(\nabla\times\mathbf{A},\nabla\times\mathbf{A}_i)-<\mathbf{B}_{\mbox{\footnotesize{self}}}\times\mathbf{n},\mathbf{A}_i>-\mu_0\left(\sigma(\dot{\mathbf{A}}+\dot{\mathbf{A}_a}+\nabla\phi),\mathbf{A}_i\right)=0,\label{Galer1}
\end{equation}
\begin{equation}
(\sigma\dot{\mathbf{A}},\nabla\phi_j)+(\sigma\dot{\mathbf{A}_a},\nabla\phi_j)+(\sigma\nabla\phi,\nabla\phi_j)-<\mathbf{n}.\mathbf{E},\phi_j>=0,\label{Galer2}
\end{equation}
where $A_i$ and $\phi_j$ are basis functions that are known \textit{a   priori}, the notation $(u,v)$ corresponds to the volume integral $\int_\Omega uv~dV$ over the volume $\Omega$, and $<u,v>$ stands for the surface integral $\int_{\partial\Omega} uv~d\mathcal{C}$. Surface terms are used for imposing Neumann boundary conditions when appropriate.

\begin{figure}[b!]
 \centering
\includegraphics[width=4cm]{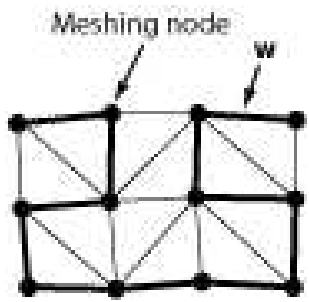}
\caption{\label{Gauge}Example of a set of meshing edges $\mathbf{w}$ used for the definition of the vector potential $\mathbf{A}$.}
\end{figure} 

The vector potential is approximated as a series of basis functions, $\mathbf{A}=\sum_ia_i\mathbf{A}_i$, where the $\mathbf{A}_i$'s are first-order edge functions that ensure the continuity of the normal component of $\mathbf{B}$ from mesh to mesh.  The vector potential is defined in a gauge for the edge functions that avoids the computation of the curl of $\mathbf{A}$ for the post-processing of the magnetic flux density~\cite{Albanese}. This gauge condition reads $\mathbf{A}.\mathbf{w}=0$, where $\mathbf{w}$ is a set of edges that connects all the nodes of the mesh through an open path, in such a way that a given pair of nodes can only be connected by a unique continuous path (see Figure~\ref{Gauge}). Similarly to $\mathbf{A}$, the scalar potential $\phi$ is expanded as $\phi=\sum_jb_j\phi_j$, where $\phi_i$ are first-order nodal functions.

\section{Computation of the finite element model}
\label{s:computation}

The weak formulation~(\ref{Galer1}-\ref{Galer2}) is implemented into the open-source solver for discrete problems, GetDP~\cite{GetDP,GetDP1}. GetDP presents two major advantages over commercial finite-element softwares: it is available free of charge and it offers a large choice of numerical methods to be implemented with a full control of the inherent parameters.

\begin{figure}[t!]
 \centering
\includegraphics[width=0.8\textwidth]{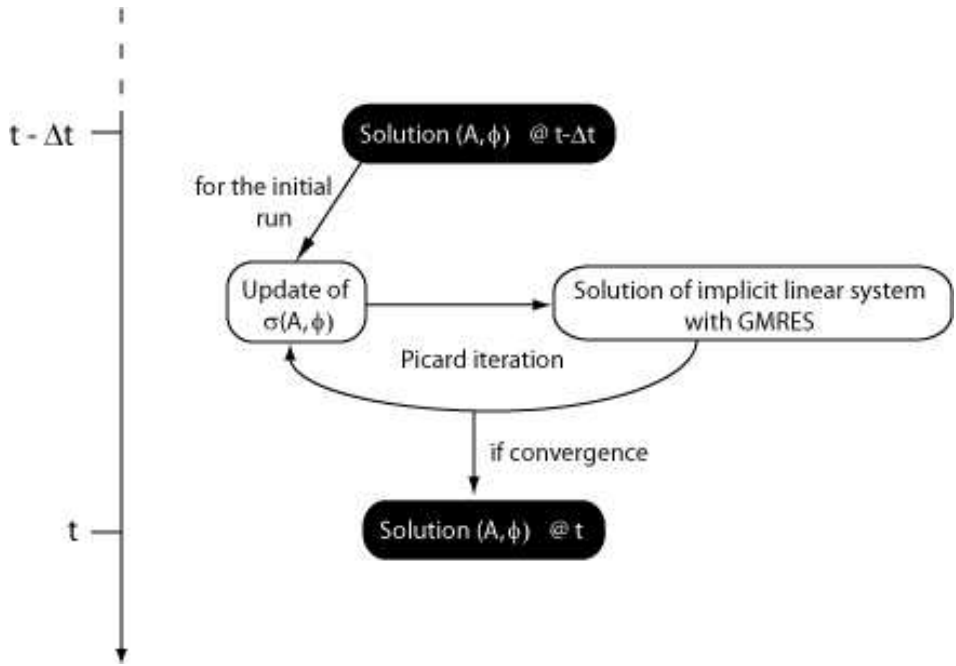}
\caption{\label{NumericScheme}Operations to be executed for one time iteration.}
\end{figure}

As stated in the Introduction, we use a step-by-step temporal resolution with a backward Euler scheme, which has a good stability and a high convergence rate even with very large time steps~\cite{NumericalMethod}. The convergence and the stability of this method has already been demonstrated in the context of HTS in the case of a $E$-formulation~\cite{Slodicka}. In our formulation, the implicit resolution required at each step generates a system of equations which are non-linear, because of the conductivity law of Equation~(\ref{E(J)}). This non-linearity is treated with a Picard iterative loop~\cite{Picard}, which consists in updating at each time step the value of the non-linear term with the solution found at the previous iteration. The loop is run until the relative difference between two consecutive solutions, $e_n$, is smaller than a predefined criterion, taken empirically here as $e_n<2.10^{-3}$. Using a Picard iteration scheme with a power law conductivity prevents one from having to deal with the infinite derivatives that appear in the more traditional Newton-Raphson scheme~\cite{Slodicka2}. Figure~\ref{NumericScheme} schematically represents the sequence of operations to be executed during a given time step.

In the following, we will compare two different choices for the time integration of the field from instant ``zero'' to a predetermined instant, $t_1$: in the first choice, the integration is carried out in a succession of small time steps of duration $\Delta t\ll t_1$; in the second choice, the equations are iterated in a single time-step, with $\Delta t=t_1$. These two choices will be compared in a number of situations.

\section{Simulation of the magnetization of a HTS tube}
\label{s:tube}

We first apply the FEM to the calculation of the magnetization of an HTS tube subjected to an axial field, in both limits of infinite and finite height. These are geometries for which the current density is everywhere perpendicular to the local magnetic field and solutions are known from other methods.  The goal of this section is to compare the FEM to these other methods to validate our approach.

The high level of symmetry in each geometry allows us in principle to reduce the mesh dimension. However, we deliberately choose not to exploit symmetry to construct the mesh, so as to use the weak formulation (\ref{Galer1})-(\ref{Galer2}) without simplification since that formulation will be used in geometries having no such symmetries. Thus, for the case of a tube of infinite extension, we use a 2D mesh of the cross section, while for the case of a tube with a finite height, we use a 3D mesh. The FEM results are compared to the predictions of the Bean model in the case of infinitely long tubes and to the results of the Green's function of Brandt~\cite{Brandt} in the case of tubes with a finite height.

\subsection{Tube of infinite extension (2D geometry)}

Consider first a superconducting tube of infinite height subjected to a uniform magnetic field applied parallel to its axis, as a ramp. The tube has an external radius $a=10~\mathrm{mm}$ and an internal radius $b=5~\mathrm{mm}$. The pinning forces in the superconductor are assumed to be infinite. Under this hypothesis, the Bean model~\cite{Bean} applies and predicts that the magnetic flux density decreases linearly inside the wall and is constant inside the hole. 

As explained in Section~\ref{s:Intro}, the power law conductivity~(\ref{E(J)}) is asymptotically equivalent to the Bean model when $n\rightarrow\infty$. From a practical point of view, it has been shown in Ref.~\cite{Cha} that the use of the power law with a critical exponent of $n=100$ and with a sweep rate of $\dot{B_a}=10~\mathrm{mT/s}$ yields an accurate approximation of the Bean model. In this section, we choose that parameter values in order to compare the results of the finite-element model to analytical expressions of the Bean critical-state. The critical current density $J_c$ is assumed to be independent on the magnetic flux density\footnotemark[1]\footnotetext[1]{Note that a model with field-dependent $J_c$ can easily be implemented in GetDP} and has a value of $J_c=2~10^7~\mathrm{A/m}^2$. The critical electric field $E_c$ is taken to be $E_c=10^{-4}~\mathrm{V/m}$. The theoretical penetration field of the tube $H_p$, is given by $H_p = J_c(b-a)=10^5~\mathrm{A/m}$, which corresponds to a flux density, $\mu_0H_p=125.6~\mathrm{mT}$.

\begin{figure}[t!]
 \centering
\includegraphics[width=0.8\textwidth]{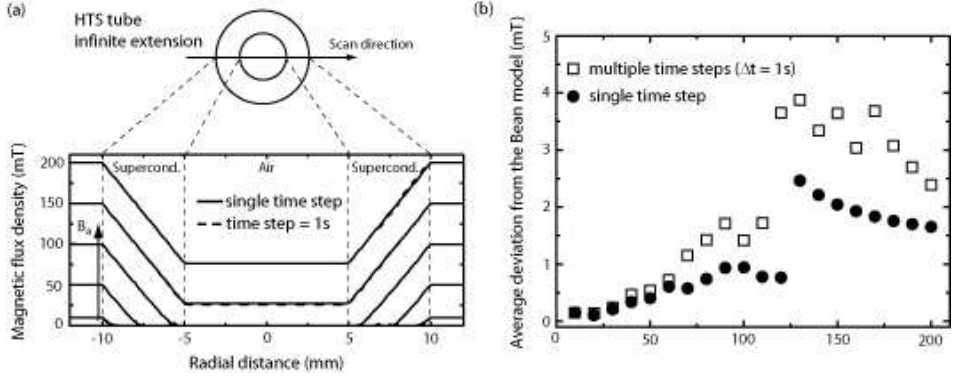}
\caption{\label{Bean4}(a)  Profile of the magnetic flux density along   the tube diameter, as calculated with the FEM with a single   time-step (solid lines) and with multiple time-steps $\Delta t = 1~\mathrm{s}$ (dashed lines).  The critical exponent $n$ is chosen large (here $n=100$) so as to approach the critical state. The magnetic flux   density is applied with a constant sweep rate of   $10~\mathrm{mT/s}$. The profiles are shown for $B_a=10,$ $50,$   $100,$ $150,$ and $200~\mathrm{mT}$. (b) Average deviation $\widetilde{\Delta B}$ from the Bean model, as a function of $B_a$, for the single time-step method (circles) and for the multiple   time-step method (squares).  }
\end{figure}

In Figure~\ref{Bean4}-(a), the magnetic field profile is plotted along the diameter of the tube for an external induction $B_a=10,~50,~100,~150,~\mbox{and}~200~\mathrm{mT}$. FEM simulations are run with different choices of time-steps: dashed lines show the results of simulations with multiple small time-steps $\Delta t = 1~\mathrm{s}$, stopping at either $t_1=1$, $5$, $10$, $15$ and $20~\mathrm{s}$; solid lines show results from single time-step simulations, where $\Delta t=t_1$ is fixed to either $1$, $5$, $10$, $15$, or $20~\mathrm{s}$. It can be observed that in each case the profile of the magnetic field in the superconductor is linear. It closely follows the result of the Bean model,
\begin{equation}
\left\{\begin{array}{lclcl} 
B_{\mathrm{Bean}} &=& B_a-\mu_0J_c (a-r) & \quad & (b\leq r\leq a)\mbox{,} \\
B_{\mathrm{Bean}} &=& B_a-\mu_0J_c(a-b) & \quad & (0<r<b)\mbox{.}
\end{array}\right.
\end{equation}
To further quantify the results, we define the average deviation from the Bean model  as
\begin{equation}
\widetilde{\Delta B}=\frac{1}{2a}\int_{-a}^{a}|B_{\mathrm{FEM}}-B_{\mathrm{Bean}}|dr,
\end{equation}
where $B_{\mathrm{FEM}}$ stands for the FEM results. Figure~\ref{Bean4}-(b) shows the average deviation in the FEM method using a single time-step (filled circles) and in that using multiple time-steps (open squares). Both methods produce almost the same deviation as long as the magnetic field has not fully penetrated the wall of the tube, or for $B_a<120~\mathrm{mT}$. For $B_a > 120~\mathrm{mT}$, the error obtained with the multiple time-step approach first increases abruptly and then has a value around $3~\mathrm{mT}$. The single time-step method, on the other hand, leads to an error which peaks at about $2.5~\mathrm{mT}$ at $B_a= 130~\mathrm{mT}$  and then decreases at larger fields to be less than $2~\mathrm{mT}$ at $B_a= 200~\mathrm{mT}$. 

Overall, the single time-step method gives an accurate solution with a relative error that stay below $5\%$ of the Bean prediction. Note that this result is obtained in a relatively short calculation time with respect to a multiple time-step method. For example, the $20$ simulations of Figure~\ref{Bean4}-(b) take less than half a day on a dual-core 2.8~GHz processor with 2~Gb of memory, whereas the multiple time-step approach with $\Delta t = 1~\mathrm{s}$ takes almost 3 days on the same computer.

\subsection{Tube of finite extension (3D geometry)}
\label{expl}
\begin{figure}[t!]
 \centering
\includegraphics[width=0.8\textwidth]{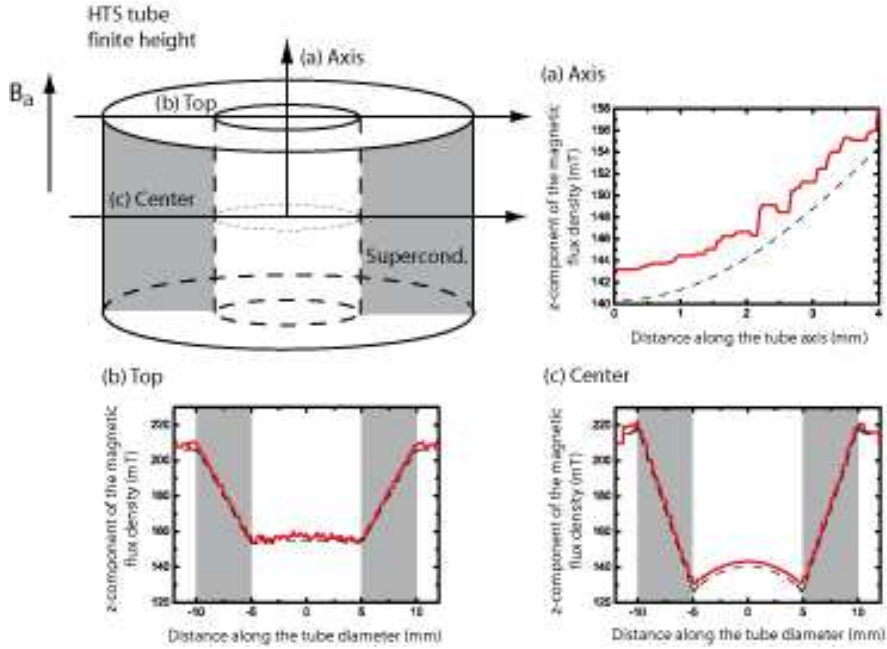}
\caption{\label{CompSam2}Sketch of the HTS tube. The outer radius is   $10~\mathrm{mm}$, the inner radius is $5~\mathrm{mm}$, and the   height is $8~\mathrm{mm}$. The arrows indicate different scan   directions for plotting the magnetic flux profile. The external   field is applied with a constant sweep rate of   $\dot{B}_a=10~\mathrm{mT/s}$, and raises up to   $B_a=200~\mathrm{mT}$.  The flux penetration problem is either   solved with the FEM single time-step method (gray solid lines and $\Delta t   = 20~\mathrm{s}$), or with Brandt's method with multiple time-steps   (black dashed lines and $\Delta t = 5~10^{-4}~\mathrm{s}$), both assuming   $n=100$.  }
\end{figure}

We now turn to the case of a superconducting tube of finite height subjected to a uniform axial field. The tube has an external radius of $10~\mathrm{mm}$, an internal radius of $5~\mathrm{mm}$ (see Figure~\ref{CompSam2}), and a height of $8~\mathrm{mm}$. The external field is applied with a constant rate $\dot{B}_a=10~\mathrm{mT/s}$ and raises up to $B_a=200~\mathrm{mT}$. Here again, a large pinning strength with $n=100$ is assumed. The critical current density, $J_c$ and the critical electric field, $E_c$ have the same values as for the tube with an infinite height.

The FEM approach is carried on a 3D mesh with a single time-step method ($\Delta t = 20~\mathrm{s}$). Only half of the tube is actually meshed, and vanishing conditions on the tangential component of $\mathrm{B}$ are imposed in the median plane. The FEM results are compared with those of the Green's function method of Brandt~\cite{Sam,Brandt}, which in this geometry is based on a 2D-kernel. The time-step is fixed at $5~10^{-4}~\mathrm{s}$ to ensure convergence.

The $z$-component of the magnetic flux density is probed along three different directions [see Figure~\ref{CompSam2}: (a), the tube axis,   (b) a diameter at the top surface, and (c), a diameter on the median plane]. Solid lines show the FEM results and dashed lines show those of Brandt's method. It can be observed that solid lines exhibit step-like features, whereas dashed lines are smoother. This difference can be traced back to the low meshing density adopted in the FEM method. Even though only half of the tube is meshed in 3D, the maximum number of available nodes with 6~Gb RAM ($200~000$) is still too small to obtain a smooth curve after linear interpolation, unlike the Brandt method which uses a specific interpolation on the 2D meshing.

Despite this observed difference, one can see that the results of the two methods are in good agreement. In particular, on the linescan along direction (c) (Figure~\ref{CompSam2}-(c)), we observe a magnetic flux density on the outer wall of the tube that is slightly larger than $B_a=200~\mathrm{mT}$. This is caused by the demagnetizing field, which was absent in the results of the previous subsection. Inside the superconducting wall, the magnetic flux density decreases linearly; in the central part of the tube, it remains at a low level, but exhibits variations due again to the demagnetizing field.

\begin{figure}[t!]
 \centering
\includegraphics[width=0.4\textwidth]{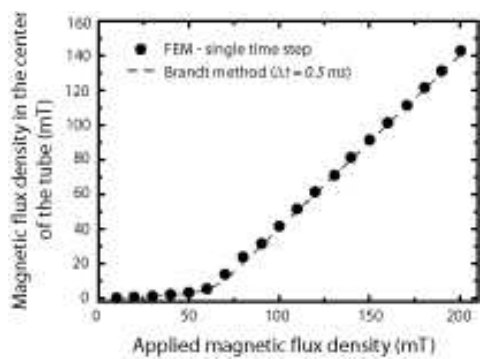}
\caption{\label{CompSam1}Magnetic flux density calculated at the center of a HTS tube with the FEM single time-step method (circles) and with the Green's function method (dashed lines and $\Delta t =   5~10^{-4}~\mathrm{s}$). The external field is applied with a ramp of   constant rate $\dot{B}_a = 10~\mathrm{mT/s}$ and increases up to $200~\mathrm{mT}$. FEM single time-step method are shown for different choices of the time-step: $\Delta t = 1,\ldots,20~\mathrm{s}$.}
\end{figure}

Figure~\ref{CompSam1} shows the $z$-component of the magnetic flux density calculated at the center of the cylinder, $B_{\mbox{\footnotesize{center}}}$, as a function of the external field $B_a$. The dashed lines show the result of the Green's function approach. Circles show the FEM results in a single time-step approach, with different choices of the time-step ranging between $\Delta t = 1~\mathrm{s}$ and $20~\mathrm{s}$. Here again, the agreement between the methods is excellent, demonstrating the relevance for adopting a single time-step iteration in a FEM approach.

\subsection{Domain of validity of the single time-step method}

\begin{figure}[b!]
 \centering
\includegraphics[width=0.4\textwidth]{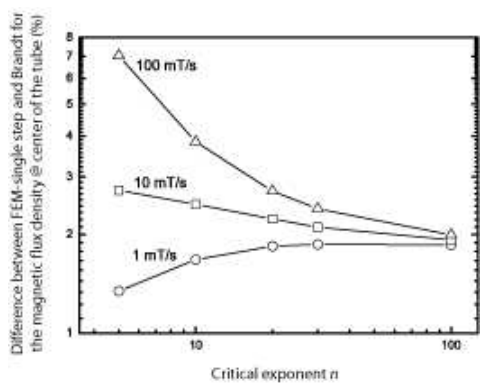}
\caption{\label{Validation}Difference between the magnetic flux density at the center of the HTS tube, as calculated by the FEM single time-step method and the Green's function method (with multiple time steps $\Delta t = 5~10^{-4}~\mathrm{s}$). The magnetic field is applied   as a ramp with a sweep rate of $1~\mathrm{mT/s}$ (squares), $10~\mathrm{mT/s}$ (circles) and $100~\mathrm{mT/s}$ (triangles). The applied magnetic flux density is ramped up to $B_a=200~\mathrm{mT}$. The critical exponent $n$ varies from $5$ to $100$. }
\end{figure}

We have seen in the two previous subsections that the FEM method with a single time-step produces accurate results in the strong pinning limit ($n = 100$). The purpose of this subsection is to analyze the accuracy of the method at lower pinning strength and establish its sensitivity to the sweep rate of the external field. Mastering these two factors is essential to make comparisons with experiments.

We estimate the error of the FEM single time-step method on the basis of the magnetic field produced at the center of the tube with a finite height (the tube considered in the previous subsection). The external field is ramped with a fixed rate $\dot{B}_a$ up to $B_a = 200~\mathrm{mT}$. The error is then evaluated as the absolute difference between the results of the FEM and the Green's function methods. Figure~\ref{Validation} shows the error (in $\%$) as a function of the critical exponent $n$, varying from $5$ to $100$, and the sweep rate $\dot{B_a}$, taken as $1~\mathrm{mT/s}$ (circles), $10~\mathrm{mT/s}$ (squares), or $100~\mathrm{mT/s}$ (triangles).  

In the very strong pinning limit ($n=100$), the error remain small (below $2~\%$) and is fairly independent of the sweep rate. This limit corresponds to the critical state which is uniquely determined by the external conditions and is independent of the magnetic history of the sample. Provided convergence is guaranteed, the FEM approach should thus produce the critical state solution. The opposite limit of low pinning strength shows a much larger sensistity to the sweep rate and a larger spread in the error. Here, these results should be considered as qualitative only, as the Green's function method itself has an error that grows in this limit, so that our estimate of the FEM error becomes questionable in this regime. For intermediate values of $n$, the error remains low and weakly sensitive to the sweep rate: e.g., for the experimentally relevant value for melt-textured YBCO at $77~\mathrm{K}$, $n=20$~\cite{Yamasaki}, the error is below $3~\%$. This demonstrates that the single time-step method is useful for simulating the magnetization of HTS with finite pinning strengths.

\section{Magnetization of drilled cylinders}
\label{s:drilled}

An extension of the single time-step method is presented in this last section where we compare the magnetization of cylinders containing 4 different arrays of holes. In a previous work~\cite{Drilled}, the Bean critical state has been used to compare the magnetization of cylinders of infinite height with four different patterns of holes: the squared and the centered rectangular lattices having a translational symmetry, and the polar squared and polar triangular lattices with a rotational symmetry. It was found that the largest trapped magnetic flux is obtained with the polar triangular lattice. We now consider FEM calculations in order to take into account demagnetisation and creep effects.

\begin{figure}[b!]
 \centering
\includegraphics[width=12cm]{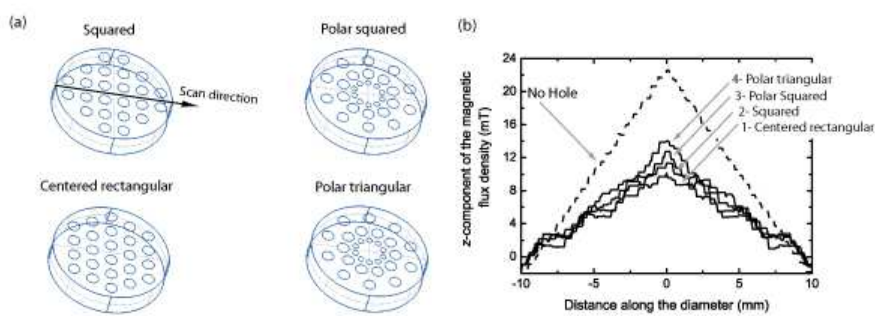}
\caption{\label{Lattices3D}(a)- The four lattices in the cylinder (radius $10~\mathrm{mm}$, height $8~\mathrm{mm}$). (b),(c) - Trapped magnetic flux profile along the cylinder diameter, as calculated with the FEM two time-steps method, with a sweep rate of $10~\mathrm{mT/s}$ and $n=25$. The profile is represented in the top cross section of the cylinder. The dashed line corresponds to the flux profile in a cylinder without holes, also calculated with the FEM two time-step method.}
\end{figure}

To this end, we consider cylinders (radius of $10~\mathrm{mm}$ and height of $8~\mathrm{mm}$) that are drilled by the four lattices considered in Ref.~\cite{Drilled}. The lattice parameters are chosen in a such way that the total diameter of the holes is constant ($50\pi~\mathrm{mm}$), so as to fix the total surface of heat exchange. The squared and the centered rectangular lattices contain each 25 holes with a radius of $1~\mathrm{mm}$. The polar lattices contain two layers of 10 holes with a radius of $1~\mathrm{mm}$ and a central layer with 10 holes with a radius of $0.5~\mathrm{mm}$. The four samples are represented in Figure~\ref{Lattices3D}-(a).

In order to calculate the trapped magnetic flux, HTS samples are first magnetized by an external field varying linearly with time. A magnetic flux is then trapped in the sample when the external magnetic field returns to $0~\mathrm{mT}$. This magnetization process is calculated here in two time-steps: one for increasing the applied magnetic flux density to $600~\mathrm{mT}$ with a constant sweep rate of $10~\mathrm{mT/s}$ and a second one for decreasing it to $0~\mathrm{mT}$ with the same sweep rate.

Figure~\ref{Lattices3D}-(b) shows the trapped magnetic flux density profile along the cylinder diameter [see black arrow in Figure~\ref{Lattices3D}-(a)]. The dashed curve corresponds to the trapped flux profile in a cylinder having the same geometry and material parameters, but containing no holes. The flux profiles exhibit steps, resulting from the low number of meshing elements used in 3D simulations as was already observed in Section~\ref{expl}. It can be observed that the maximum trapped magnetic flux density is smaller in the drilled samples than in the bulk one.

Table~\ref{MaxTrapped} lists the values of the maximum trapped magnetic flux density in the top cross section and in the median plane, as well as the results obtained for infinite   cylinders~\cite{Drilled}. In all cases, the maximum trapped magnetic flux density is obtained with a polar triangular lattice, with a value higher by $\approx~40\%$ with respect to that obtained in a centered rectangular lattice. This result is independent of the cross-section where it is calculated and agrees with the theoretical predictions based on the Bean model~\cite{Drilled}. The demagnetisation effects only affect the values of the maximum trapped flux density that are smaller in the finite height samples than in the cylinders of infinite height with the same hole lattice.

\begin{table}[t]
\centering
\begin{tabular}{rccc}
 \hline\hline
$B_{\mathrm{max}}$ (mT)&\scshape{3D - top}&\scshape{3D - center}&\scshape{Infinite height}\\
\hline
Polar triangular lattice &   70.05 &112.7&137.9 \\
Polar squared lattice &   63.6&97.7&120.9\\
Squared lattice &  56.7&87.8 &110.8 \\
Centered rectangular lattice & 50.6&76.4&101  \\
\hline \hline
\end{tabular}
\caption{\label{MaxTrapped}Maximum trapped magnetic flux density in cylinders of finite height, as calculated in the top cross section (3D - top) and in the median plane (3D - center), and in cylinders of infinite height obtained in Ref.~\cite{Drilled}. }
\end{table}

\section{Conclusions}
\label{s:Conclusion}

Using the \textit{open source} solver GetDP, we have implemented a 3D finite element $A$-$\phi$ formulation for the calculation of the magnetization of bulk HTS subjected to a ramp of magnetic field. The numerical method is based on a single time-step iteration that reduces drastically the total calculation time. By comparing it to the Bean model~\cite{Bean} in infinite tubes and to the Green's function method~\cite{Brandt} in tubes of finite height, we have shown that the FEM approach accurately describes the magnetic properties of superconductors with strong pinning. Although it neglects the effects associated with currents that are parallel to the magnetic field, that study makes progress toward a 3D model of HTS magnets that takes into account demagnetisation effects and flux creep.
 
As an extension of the FEM single time-step method, we have calculated the trapped magnetic flux in drilled cylinders of finite height. The numerical method uses only two time-steps: the first one during the ramping up of the applied field to $H_{\mathrm{max}}$ and the second one for the return of the external field to zero.  Using this method, we have been able to extend a previous analysis for tubes of infinite extension to a full 3D geometry. These results confirm that the trapped magnetic flux is maximized by drilling the holes according to a polar triangular lattice.

\section*{References}

\bibliographystyle{unsrt}
\bibliography{FEM_HTS_Lousberg}

\end{document}